\documentclass[useAMS,usenatbib]{mn2e}
\usepackage{graphics}
\usepackage{subfigure}
\usepackage{graphicx}
\usepackage{amssymb}
\usepackage{hyperref}
\title
[Cosmic queuing: Satellites and Building Blocks]
{Cosmic queuing: galaxy satellites, building blocks and the hierarchical clustering paradigm}
\author[Claudia del P. Lagos, Nelson D. 
Padilla, Sof\'ia A. Cora]{Claudia del P. Lagos$^{1}$, Nelson D.
Padilla$^{1}$, Sof\'ia A. Cora$^{2,3}$\\
$^{1}$Departamento Astronom\'ia y Astrof\'isica, Pontificia Universidad
Cat\'olica de Chile, Av. Vicu\~na Mackenna 4860, Stgo., Chile\\
$^{2}$ Facultad de Ciencias Astron\'omicas y Geof\'isicas de la Universidad Nacional
de La Plata, and Instituto de Astrof\'isica de La Plata \\(CCT La Plata, CONICET, UNLP), Observatorio Astron\'omico, Paseo del Bosque S/N, 1900 La Plata, Argentina\\
$^{3}$Consejo Nacional de Investigaciones Cient\'{\i}ficas y T\'ecnicas,
Rivadavia 1917, Buenos Aires, Argentina\\
}
\begin{document}

\date{Accepted ???. Received ???; in original form ???}

\pagerange{\pageref{firstpage}--\pageref{lastpage}} \pubyear{2009}

\maketitle

\label{firstpage}

\begin{abstract}
We study the properties of building blocks (BBs, i.e. accreted satellites) and surviving
satellites of present-day galaxies using the {\small SAG} semi-analytic model 
of galaxy formation in the context of a concordance $\Lambda$ Cold Dark Matter 
($\Lambda$CDM) cosmology.  
{We consider large numbers of DM halo merger trees spanning a wide 
range of masses ($\sim 1 \times10^{10} - 2.14\times10^{15} \, M_{\odot}$).} 
We find higher metallicities for BBs {with respect to surviving satellites}, 
an effect produced by the same processes behind the build-up of the mass-metallicity relation.  
{We prove that these metallicity differences arise from the higher peak
height in the density}
fluctuation field occupied by BBs 
and central galaxies which have collapsed into a single object earlier than 
surviving satellites. 
BBs start to form stars 
earlier, during the peak of the merger
activity in $\Lambda$CDM, and build-up half of their final stellar mass (measured 
at the moment of disruption) up to four times faster than surviving satellites.  
Surviving satellites 
keep increasing their stellar masses rather quiescently down to $z\simeq1$.  
The difference between the metallicities of satellites, 
BBs and central galaxies depends on the host DM halo mass, 
in a way that can be used as a further test for the concordance cosmology.
\end{abstract}

\begin{keywords}
galaxies: formation - galaxies: evolution 
\end{keywords}

\section{Introduction}\label{Introsec}

Recent times have witnessed a growing discussion regarding the formation of
present-day galaxies via the complete disruption of satellites by more massive central
objects (e.g. as proposed by \citealt{Searle78}, SZ78 hereafter),  
in line with the prediction from the Cold Dark-Matter (CDM) model of a bottom-up 
scenario of structure formation.  
{Those} disrupted satellites are
usually referred to as building blocks (BBs; \citealt{Ibata94}).  

The observational study of the Galactic halo by SZ78
provided hints pointing to a hierarchical galaxy formation scenario.  
They studied metallicities of the outer Milky Way (MW)
{stellar}
 halo and inferred that it was formed
from infall of small protogalactic fragments characterised by 
chemical evolution processes similar to
the present-day dwarf spheroidal (dSph) MW satellites. 
Although a hierarchical scenario implies, to some degree,
the infall of BBs as suggested by SZ78,
it does not imply that BBs are characterised by formation histories {(i.e.
star formation rate evolution)} similar to those of
present-day satellites, as has been suggested by 
Geisler et al. (2007, G07 hereafter).

Several observational facts provide indications that BBs
and satellites should be characterised by different formation histories,
since the galaxy population seems to have evolved with time as shown by
(i) measurements of the evolution of the stellar mass
function up to $z \approx 5$ (Drory et al. 2005), and of the luminosity
function (e.g. \citealt{Faber07}),
(ii) the decline of the star formation rate towards $z=0$ (see for instance
\citealt{Hopkins05}), 
(iii) the important increase in the frequency of galaxy mergers 
at high redshift,
observed as violent starbursts (e.g. \citealt{Fontana04}), and
(iv) 
{the differences in galaxy formation epochs, where massive galaxies appear to be older than low-mass
objects (e.g. \citealt{Cowie96})}.
All these arguments indicate that galaxies at high redshift are likely to be 
different than
their present-day counterparts and, therefore, 
BBs are also probably different from
present-day satellites.  
{This is also supported by several
studies of our local group
which indicate that 
Galactic halo stars, thought to come from remnants of previous
accretion events, are characterised by
different abundance patterns than present-day dSph galaxies.
Namely, extremely iron poor stars ($[{\rm Fe}/{\rm H}]<-3$) 
are not found in dSphs \citep{Helmi06},} 
and typical `accreted' halo stars show $\alpha$-elements to iron ratios
$\sim 0.1-0.3$ dex higher than stars in dSphs (see G07).

On the theoretical side, 
the galaxy population in the currently favoured hierarchical cosmology, obtained via either
semi-analytic models or gas dynamic simulations, is also
observed to evolve in broad agreement with observations (e.g.
\citealt{DeLucia06}; \citealt{Libeskind06}; \citealt{Lagos08}).
This galaxy evolution is mainly driven by a changing frequency of DM halo mergers which
shows a peak activity at redshifts $z\simeq 2-3$ \citep{Okamoto05}.
The particular case of the MW has been studied in a $\Lambda$CDM scenario 
(\citealt{Robertson05}; 
\citealt{Bullock05}; 
\citealt{Font06a}, \citealt{Font06b}; 
\citealt{DeLucia08}) from which it arises that
surviving satellites and BBs are characterised by different metallicities,
as a result of differences in their formation time-scales. 

In this work, we investigate this issue using a considerably larger number of MW type haloes (a total of $142$).
Additionally, with the aim to obtain clues about formation scenarios at different mass scales, we also
extend the study of building blocks and central/satellite galaxy populations
to a wide range of DM halo masses.  We focus on the systematic differences 
in metallicity (i.e. $\rm log_{10}(Z/Z_{\odot})$), 
age and formation time-scales
between central galaxies (those hosted by the largest sub-halo in a DM halo),
satellites (all other present-day galaxies) and BBs (galaxies that have 
already merged with a central galaxy).
Taking into account the assumption of SZ78, we consider that, as
a result of tidal effects, a fraction of the stellar component of the BBs
contribute to form the stellar halo of the central galaxy instead
of converging to its bulge.

We use the semi-analytic model of galaxy formation
{\small SAG} (acronym for `Semi-Analytic Galaxies')
by Lagos, Cora \& Padilla (2008, LCP08 hereafter), in combination with a
$\Lambda$CDM {\em N}-body simulation characterised by a periodic box of $60
\,$h$^{-1}\,{\rm Mpc}$, with a resolution of $1.001 \times
10^{9}\,$h$^{-1}\,{\rm M}_{\odot}$ per DM particle. The simulation parameters
are consistent with the results from WMAP data 
(\citealt{Sanchez06}, $\Omega_{\rm m}=0.28$, $\Omega_{\Lambda}=0.72$ and
$\sigma_{8}=0.9$; a Hubble constant 
$H_{0}=100 \, h\, {\rm Mpc}^{-1}$, with $h=0.72$).
As a result of the median-size of our 
comoving box, the maximum halo mass is $M=5.34\times10^{14} M_{\odot}$.  We complement our
study by using the non-radiative {\em N}-body/SPH (Smoothed Particle Hydrodynamics) 
simulations of galaxy clusters of masses
$M\sim2.14\times10^{15} M_{\odot}$ \citep{Dolag05} used in 
\cite{Cora08},
extending our dynamical range and, consequently, 
providing a better sampling of the
high-mass end of the galaxy population.  
{Because of resolution constraints,
our analysis is restricted to galaxies with stellar masses
$M_{\rm Stellar} > 10^7 M_{\odot}$; 
i.e. none of our conclusions refer to very low mass objects such as the recently
discovered ultra faint
dwarfs (e.g. \citealt{Simon07}).}
In order to distinguish results intrinsic to
the $\Lambda$CDM scenario, we also use a version of the SAG model 
where disc instabilities and feedback from active galactic nuclei (AGN) 
are {switched off}
{(maintaining the parameters of the other physical processes fixed,
i.e, star formation, supernovae (SNe) feedback).}
In the remainder of this letter, we will refer 
to {\small SAG} and its modified version as models A and B, respectively. 
Finally, as we are interested in the properties of the
MW and its satellites,
we mimic such a population by selecting
$z=0$ galaxies with circular velocities in the range
$v_c=210-230\,{\rm km}\,{\rm s^{-1}}$.  Such galaxies
{are hosted by DM haloes of masses}
$ 10^{12} M_{\odot} \la M_{\rm
DM} \la 4\times 10^{12} M_{\odot}$.  In the remainder of this letter, the $142$ haloes selected this
way will be referred to
as MW-type haloes.

\section{Building blocks of present-day galaxies}\label{building}

\begin{figure}
\begin{center}
\vskip -.3cm
\includegraphics[width=0.39\textwidth]{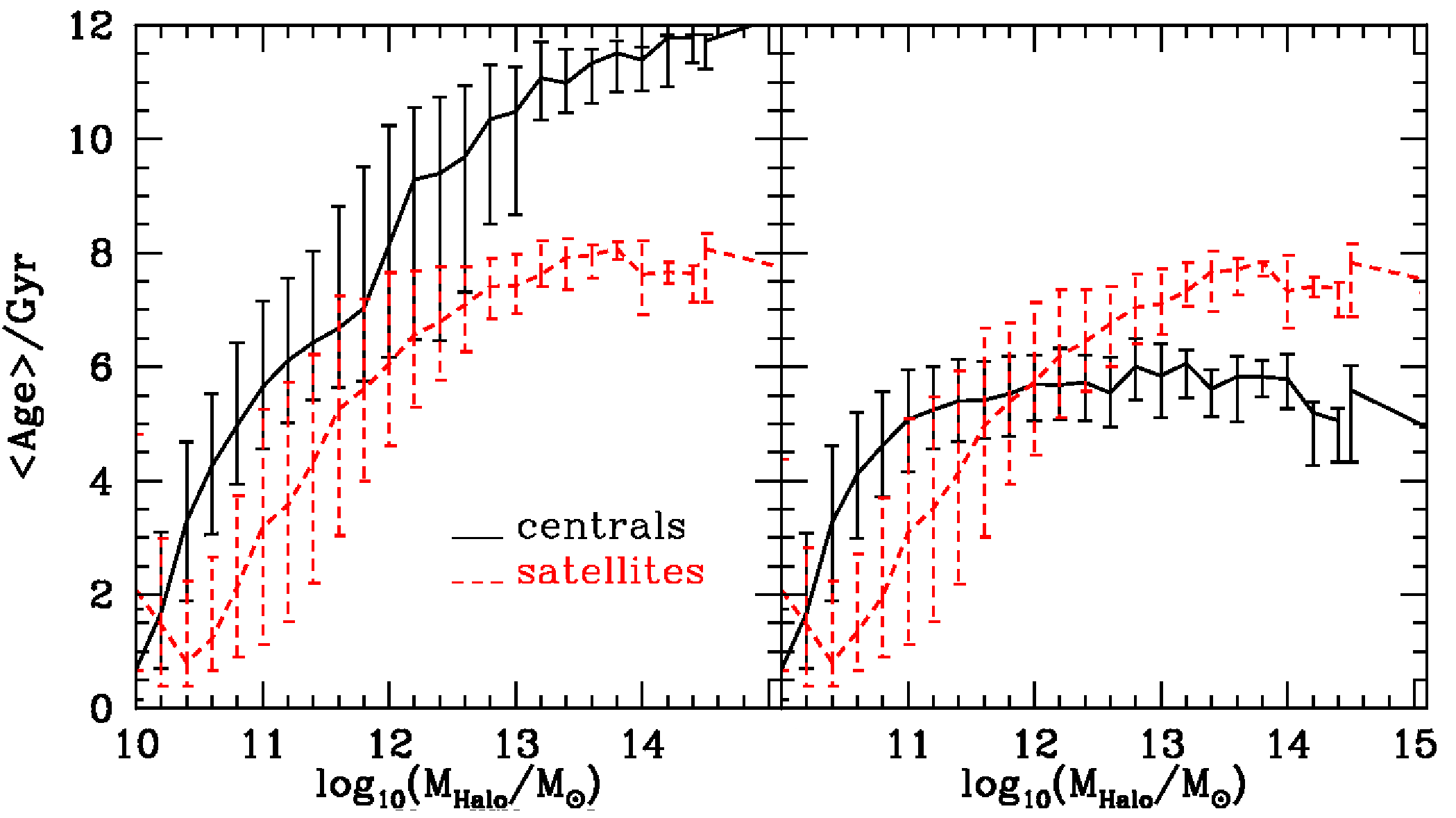}
\includegraphics[width=0.39\textwidth]{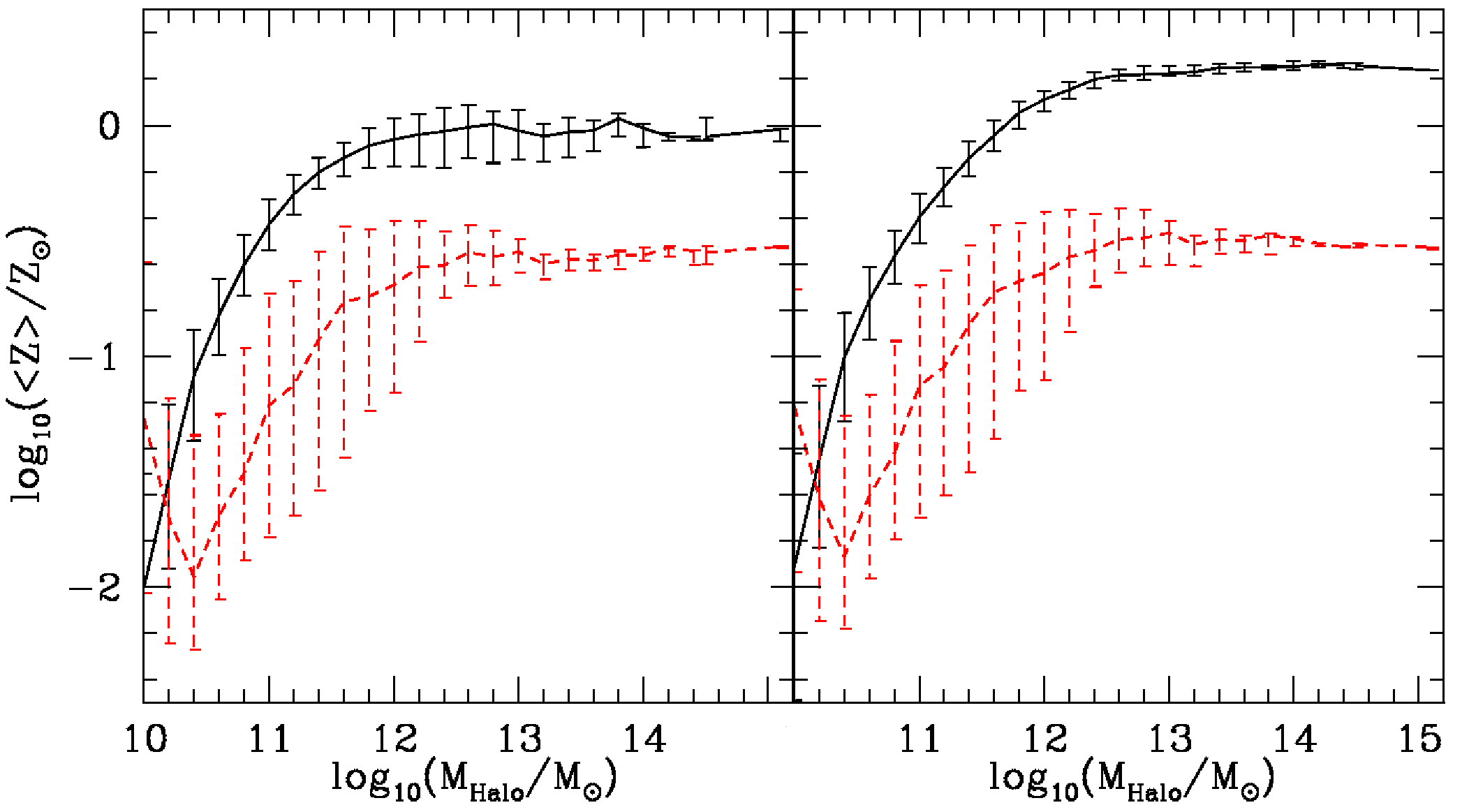}
\caption{Average age and metallicity for the satellite galaxy population 
(dashed
lines) 
and central galaxies
(solid lines), 
as a function of 
their host DM halo mass. Results for models A and B are shown in the 
left and right panels, respectively. Errorbars correspond to the $20$ and $80$ percentiles.}
\label{MZ}
\end{center}
\vskip -.5cm
\end{figure}

\begin{figure}
\begin{center}
\vskip -.3cm
\includegraphics[width=0.3\textwidth]{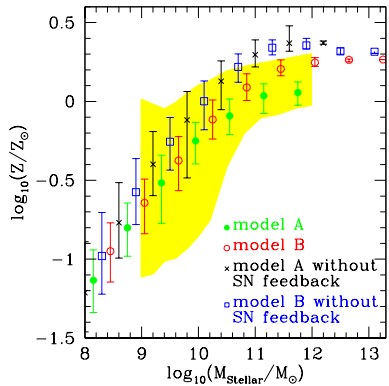}
\caption{Stellar mass-metallicity relation at $z=0$ of the full galaxy 
population
for model A,
model B, 
and model A and B without
SNe energy feedback (symbols are indicated in the figure key).
Errorbars correspond to the $10$ an $90$ percentiles.
 The observational relation determined by
\citet{Gallazzi05} is represented by the shaded area.
}
\label{SN}
\end{center}
\vskip -.5cm
\end{figure}

\vskip -.1cm
As a first step, we analyse the way in which
galaxy metallicities\footnote{The
detailed implementation of metal enrichment in {\small SAG} (i.e. yields
from
core collapse SNe {(SNe CC)}, SNe type Ia {(SNe Ia)}
and low-intermediate mass stars) is described in \citet{Cora06}.}
and stellar ages depend on
the host DM halo mass; stellar ages are calculated using the stellar mass
weighted mean.
Fig.~\ref{MZ} shows the results for models A (left panels) and B 
(right panels), where dashed and solid lines represent,
respectively, the average properties of satellite and central galaxies for each DM halo.
Model A shows a shift in galaxy ages between central and satellite galaxies 
(upper left panel), where the latter are usually at least $\sim 1-2$~Gyr younger.
The magnitude of this shift depends on the DM halo mass, where more massive 
haloes show larger age differences.  These are a consequence of the AGN feedback, 
which reduces or even stops gas cooling in massive central galaxies at low redshifts, thus
improving the agreement of galaxy luminosities and colors with observational results 
(see LCP08). In model B, central galaxies are younger than their
satellites for DM halo masses $M_{\rm DM} \ga 10^{12} M_{\odot}$ as a result
of the absence of this heating source. 
Both models show a clear correlation between galaxy metallicity and DM halo mass 
(lower panels of Fig.~\ref{MZ}), with central galaxies being more chemically 
enriched than their surviving satellites.
AGN feedback is not decisive for the build up of the relation between metallicity and halo mass
for both central and satellite galaxies, as becomes evident from model B.
This tight relationship is a direct consequence of the hierarchical clustering process, 
where more massive galaxies sit in initially higher density peaks and, as a consequence,
start to form earlier on average \footnote{Assuming a linear evolution for density
fluctuations proportional to the scale-length of the Universe, $a$,
and non-linear fluctuation for collapsed objects $\propto a^3$.},
allowing them to acquire higher amounts of metals.
Thus, a hierarchical scenario is prone to produce less chemically
enriched satellites with respect to their central galaxies, regardless of the
AGN feedback modeling
(also suggested by Tissera, De Rossi \& Scannapieco, 2005).
{Note that
the differences in metallicity between central and satellite galaxies become
larger as the halo mass increases. 
{In the case of MW-type haloes}
this difference is on average $\approx 0.8$ dex, and can be as high as $\approx 1.3$ dex.

The different metallicities of central and satellite galaxies arise from the 
combination of the {stellar} mass-metallicity relation {(SMMR)},
and the lower increase of the median stellar mass of satellites
compared with that of their central galaxy as the
mass of the DM halo increases (e.g. stellar masses
for satellites are $1.03$ dex lower for 
$M_{\rm Halo} \approx 3\times10^{11} M_{\odot}$,
and $1.35$ dex lower for $M_{\rm Halo} \approx 10^{12} M_{\odot}$, on average).}
Figure \ref{SN} shows the {SMMR}
for the 
{full} galaxy population in model A 
(filled circles), model B (open circles), 
{and model A and B without SNe energy feedback (crosses and open squares,
respectively).}
As can be seen,
SNe feedback is essential to provide a good agreement with the
observed data (shaded area, \citealt{Gallazzi05}); neglecting to include it
produces significantly higher metallicities.
However, the increasing trend of the {SMMR}
is obtained 
independently of the details of the 
heating sources that affect the evolution of the baryonic component, including SNe energy feedback.
The SMMR of the individual populations of central and satellite galaxies 
follow closely that of the global population,
with only minor differences at
the low mass end.

Notice that the previous metallicities correspond to the full stellar content of
galaxies. Therefore, we cannot directly compare these results to
measurements of metallicities of the MW stellar halo.
{As was mentioned above, we take the BBs at the time of the merger as a
representative population of the stellar halo.}
This assumption imposes that, in order to be able to compare the properties of both populations, 
our analysis includes all simulated galaxies which have experienced 
at least one merger (with a BB) during their lifetime and have at least one
surviving satellite.

\begin{figure}
\begin{center}
\vskip -.4cm
\includegraphics[width=0.3\textwidth]{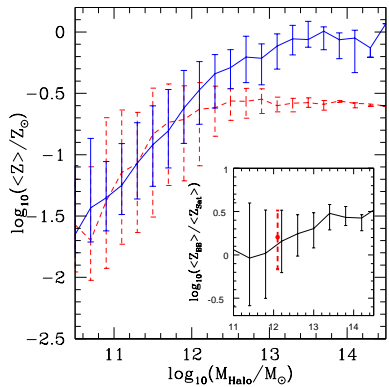}
\caption{
Average metallicity for the present day satellite galaxy population in individual
DM haloes (dashed line) and for the BBs of central galaxies (solid line), as a function of 
the host DM halo mass, for model A.  
Errorbars correspond to the $20$ and $80$ percentiles.
The inset shows the difference between the average metallicities
of BBs and satellites
as a function of DM halo mass. The filled circle represents
MW-type haloes. 
}
\label{MZ2}
\end{center}
\end{figure}

A first comparison shows that the stellar masses of BBs 
(just before the accretion) are only slightly lower than that of $z=0$ surviving satellites,
and that BBs have slightly higher $M_{\rm ColdGas}/M_{\rm Stellar}$ fractions (differences
of a factor of $\sim 2$ for MW-type haloes).
Fig.~\ref{MZ2} shows, {for model A}, the average metallicity of surviving
satellites (dashed line) and BBs (solid line) 
as a function of the host DM halo mass. Important 
differences between both populations arise for DM halo masses
$M_{\rm DM} \gtrsim 6 \times 10^{11} M_{\odot}$; at lower values, the 
metallicities between satellites and BBs are rather similar, as a result of 
the smaller differences between formation epochs 
for BBs and satellites of galaxies in lower 
density peaks. 
{We define the formation epoch as the time when a galaxy has acquired 
10\% of its final stellar mass, thus avoiding ambiguities arising from the 
resolution limit.}
The inset shows the difference between average 
metallicities of BBs and satellites, $\rm log_{10}(\langle Z_{\rm BB}\rangle/\langle Z_{\rm Sat}\rangle)$, 
in individual systems; the filled circle shows MW-type haloes for which
the difference in the average metallicities is $\approx 0.2$ dex.  
This measurement is consistent with 
results for the MW, where the abundance patterns shown by dSphs and the galactic halo 
are suggestive of higher metallicities for the MW BBs (e.g. \citealt{Tolstoy04}). 
In our sample, $70$ per cent of the MW-type haloes
show $\rm log_{10}(\langle Z_{\rm BB}\rangle / \langle Z_{\rm Sat} \rangle ) \ga 0.0$, 
which indicates that the MW stellar content
would be rather typical for its host halo mass. 
For model B, we find a very similar behaviour, indicating that the
difference between metallicities of BBs and surviving satellites 
is an intrinsic feature 
of the $\Lambda$CDM scenario. 
This 
claim is also supported by the similarity of our results for
different dynamical friction time-scales; we 
use the two alternative recipes implemented in SAG, as described in LCP08.
{Our results confirm those}
found for MW-type haloes in
numerical simulations (e.g. \citealt{Robertson05}), but this
is the first time such an analysis is performed using an unbiased sample of merger trees 
(with at least one galaxy merger and one surviving satellite) 
extracted from a fully non-linear cosmological numerical simulation.

It is important to remark that our model predicts that the majority of the stars in 
galaxies hosted by MW-type haloes are formed in-situ since BBs only
contribute with $\approx 15$ per cent of the $z=0$ stellar mass. 
In contrast, more than $50$ per cent of the $z=0$ stellar mass
of central galaxies hosted by DM haloes with masses $M_{\rm DM} \ga 10^{13} M_{\odot}$ 
comes from the accretion of BBs.

\begin{figure}
\begin{center}
\vskip -.8cm
\includegraphics[width=0.47\textwidth]{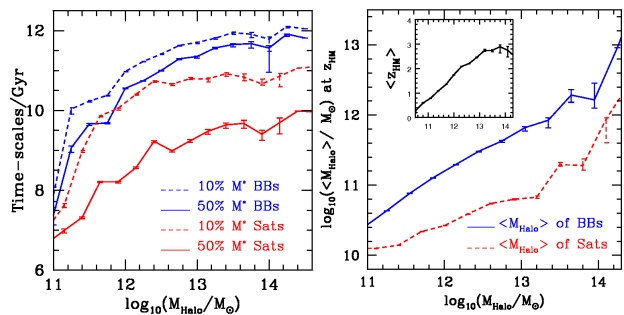}
\caption{\textit{Left panel:} Average LBT to the formation
of a $10$ 
and $50$ per cent 
of the final stellar mass of 
BBs 
and surviving satellites 
as a function of DM halo mass (line types are indicated in the figure key), for
model A. 
\textit{Right panel:} Central halo mass at the redshift of merger with the parent
haloes of BBs 
and the average mass of parent haloes of
surviving satellite galaxies at the same redshift.
The average redshift for the merger is shown in
the inset. Errorbars show the uncertainty on the mean. 
}
\label{TS}
\end{center}
\end{figure}

By analysing 
the details of their formation {(i.e. formation epoch and history)}, we can explain the gap in metallicity between 
BBs and surviving satellite galaxies.  We estimate their formation time-scales
by taking into account the look-back time {(LBT)} 
to the moment in which they acquire $10$ 
and $50$ per cent of their final stellar mass. These are represented {for
model A} in the left panel of Fig. \ref{TS}
by dashed and solid lines, respectively, for BBs (blue) and surviving satellite galaxies (red),
as a function of the host DM halo mass.
As can be seen, regardless of environment, 
BBs assemble their stellar mass in considerably shorter times
($\sim 0.25$~Gyr to increase
their stellar mass from a $10$ to a $50$ per cent of its final value) than surviving
satellites ($\sim1.5$~Gyr), 
implying that the former will naturally show enhanced $\alpha$ element 
abundances in comparison to the latter.
{During short and intense star formation (SF) events, such as those characteristic of BBs,
a stellar population achieves
large $[\alpha/{\rm Fe}]$ ratios 
since stars are formed from gas polluted by SNe CC products;
the onset of SNe Ia
occurs $\sim 1$~Gyr after the first SF episode,
when the bulk of SF activity has already occurred.}
{A detailed study of 
$\alpha$ element abundances will be tackled in
a forthcoming paper.} 
Additionally, BBs also finish acquiring $50$ per cent of their final 
stellar mass substantially earlier ($\sim 2$~Gyr) than surviving satellites
as a result of their different initial environments.  
BBs are aggregated into more massive
objects earlier than surviving satellites; i.e. their
initial location could be part of a higher density fluctuation than that
of satellites and, therefore, collapse earlier, 
as already discussed
in the literature (e.g. \citealt{Helmi06}).
In model B, we consistently find that BBs form both earlier and faster than 
surviving satellites regardless of their host DM halo mass,
indicating that these two effects are natural in a hierarchical universe.

These results indicate that {the formation epoch of BBs in MW-type haloes corresponds to
$z\simeq2.7$ 
(with $\simeq 8$ per cent
dispersion in LBT in our model)
and that they acquire $50$ per cent of their stellar content 
by $z\simeq2.25$,
whereas MW satellites show formation epochs at $z\simeq2.1$ (with 
$\simeq 5$ per cent dispersion 
in LBT
for MW-type
haloes) and acquire $50$ per cent of their stellar mass by $z\simeq1.25$. }
In the concordance cosmology, the peak in the merger activity 
of galaxies occurs at $z\approx 2-3$ (e.g. LCP08),
which coincides with the formation epoch of BBs, and only partially overlaps the
epoch when surviving satellites start acquiring their stellar mass.  
Therefore, the cosmological peak 
merger activity, {which is directly linked to the maximum SF activity in the
Universe}, may be partly responsible for the different properties of BBs
and satellites and would explain the fact that their properties are more
similar at lower {host DM halo} masses {(see Fig.~\ref{MZ2})}, 
for which most of the SF activity 
occurs at later times
{when the cosmic SF activity is already in its declining phase.}
Note that at fixed DM halo mass, the populations of BBs on the one hand,
and of surviving satellites on the other, show very homogeneous properties;
the relative 1-$\sigma$ dispersion in ages and
metallicities is of the order of $10$ per cent, or smaller.

The right panel of Fig.~\ref{TS} shows {in} a solid line the mass of the 
{central} DM halo
at the moment of the accretion of the parent DM haloes of BBs 
(when the collapse of the
density peak has just occurred and before dynamical friction starts the 
merging process between the BBs and the central galaxy) 
as a function of the $z=0$ host
halo mass.
The average DM halo mass hosting
the $z=0$ surviving satellites at the same {LBT} is shown by a dashed line.
Additionally, the average redshift of acquisition of the host haloes of BBs
is shown in the inset. 
We can see that, at the accretion epoch of BBs and 
regardless of environment, the mass of haloes hosting BBs is larger than that of the
surviving satellites.
This result {proves} that BBs are embedded in higher initial 
fluctuation peaks than surviving satellites.
This way, a cosmic queuing picture emerges 
where the progenitors of the most massive present-day objects start their formation process 
earlier than lower mass 
systems.
Lower present-day mass objects need to wait longer to reach this collapse threshold 
(see, for instance, \citealt{Press74}).

In summary, the $\Lambda$CDM cosmology predicts that
central and satellite galaxies represent two distinct populations with different  age and
metallicity distributions. Even though satellite galaxies in general cannot be expected to resemble the
original galactic building blocks, they can still be used to probe the assumptions
behind galaxy formation models by comparing their metal 
content to that of the halo stars of their central galaxy.  

\vspace{-0.5cm}
\subsection{Why are all the MW satellites so similar?}

The observational result that MW satellites are similar between themselves 
in terms of 
metallicity properties (G07), 
and different from their host -- the MW galaxy -- has
been a subject of intense debate. In this letter, we suggest the latter is a consequence of the same
processes behind the build-up of the {SMMR}; regarding the former,
we find that the dispersion in {the stellar mass and age for satellites
seldom exceeds $23$ and $10$ per cent, respectively. In particular, MW-type haloes
show dispersions below $5$ per cent for the formation {LBT} and $20$
per cent for the stellar mass of 
surviving satellites.} Such narrow
spreads in masses and ages also results in narrow metallicity dispersions within 
DM haloes, providing a plausible
explanation for the similarities between the different MW dSphs.

\section{Summary and conclusions}\label{conclusion}

In this work, we have used the semi-analytic model described by LCP08 in 
combination with a cosmological {\em N}-body simulation of the concordance $\Lambda$CDM 
cosmology and 
hydrodynamical simulations of galaxy clusters,
to study the populations of surviving satellites and building blocks 
of central galaxies hosted by DM haloes of various masses (model A).
In order to isolate the effects arising from the hierarchical growth of structures alone,
we also considered a model without AGN feedback 
and disc
instabilities (model B).

Our main results are summarised as follows.
\newline (i) On average, massive DM haloes host 
more chemically 
enriched galaxies than low-mass haloes, {regardless of the inclusion of AGN
or SNe feedback.}
In model A
central galaxies are older 
than the surviving satellite population for all the halo masses considered.
In model B, central galaxies 
{of} the most massive haloes can be younger than their
satellites.
The latter effect arises as a result of the lack of the AGN feedback,
which suppresses the star formation in massive galaxies 
{at low redshift}. 
\newline (ii) Halo stars (analysed via the stellar content of BBs) 
and present-day satellite galaxies show clear differences
in age and metallicity; the former are found to be more chemically enriched than the 
latter, regardless of the inclusion of energy feedback sources as AGN and SNe.
We find that these results 
arise 
from the earlier 
formation epochs
and faster formation 
time-scales experienced by BBs with respect to satellites. The epoch of 
assembly of BBs for MW-type galaxies roughly corresponds to
the peak merger activity in a $\Lambda$CDM universe.
\newline (iii) With respect to MW-type DM haloes in {\small SAG}, 
we find that, in 
$70$ per cent of the cases,
the metallicities of BBs are higher than those of surviving satellites, as
is thought to be the case of the MW (e.g. G07).
{This indicates that the MW
stellar content would be rather typical for its host DM halo mass.} 
This particular
result does not depend on the implementation of AGN feedback and disc instabilities.
However, it should be borne in mind that 
the observational data may still be {subject} to selection biases as has been proposed by
Bullock \& Johnston (2005). 

Our conclusions do not depend on the detailed physical assumptions adopted in the
semi-analytic model but are rather a natural consequence of the hierarchical
growth of structures.
To conclude, we remark that the differences between BBs and surviving 
satellites can be thought of as a consequence of the existence of a {SMMR},
which is found both in observations and in models of galaxy formation. This
relation indicates that objects with higher stellar masses, corresponding to higher peaks
in the density field, are characterised by higher metallicities. Therefore, it is
not surprising that central galaxies and BBs (which collapsed early into
massive objects and populate high peaks) are characterised by higher metallicities than satellite galaxies.
The latter only recently began their merging process with 
their host galaxy
and come from lower peaks in the density fluctuation field.
This 
explains their lower metal
abundances.  
In this picture, satellite galaxies are simply waiting their turn to become BBs
to imprint their signature in the continuously evolving relation between the properties of surviving
satellites and their host galaxy stellar halo.

\section{Acknowledgements}
\vskip -.1cm
We thank Carlton Baugh, Manuela Zoccali, Doug Geisler and
Amina Helmi for helpful discussions and comments. 
We also thank Klaus Dolag and Roberto Gonz\'alez for providing the
cosmological simulations.
We acknowledge the anonymous referee for
useful remarks which helped improve this work.
NP was supported by Fondecyt grant No. 1071006.  The
authors benefited from a visit of SC and Carlton Baugh to Santiago de Chile supported by
Fondecyt grant No. 7080131. This work was supported in part by the 
FONDAP Centro de Astrof\'isica, by Basal project PFG0609,
and by
PIP 5000/2005 from Consejo Nacional de Investigaciones
Cient\'ificas y T\'ecnicas, Argentina.

\label{lastpage}

\end{document}